\documentclass{research4cacm}

\usepackage{times}
\usepackage{graphicx}
\usepackage[hyphens]{url}
\usepackage[hyperindex,breaklinks]{hyperref}
\usepackage[shortcuts]{extdash}

\long\def\com#1{}

\newcommand{\xxx}[2][XXX]{}

\newcommand{\baf}[1]{\xxx[BAF]{#1}}

\date{}

\newcommand{\eg}{{\it e.g.}}
\newcommand{\ie}{{\it i.e.}}

\newcommand{\enc}[2]{\textsc{ENC}_{#1}(#2)}

\title{Seeking Anonymity in an Internet Panopticon \\
	{\tt UNPUBLISHED DRAFT}\thanks{This material 
is based upon work supported by the Defense Advanced Research Projects Agency 
(DARPA) and SPAWAR Systems Center Pacific, Contract No.~N66001-11-C-4018.}
}

\author{
	Joan Feigenbaum and Bryan Ford \\
	Yale University}

\begin{document}

\maketitle

\baf{Work in somewhere: anonymity attacks and defenses as an
	economic cost/benefit system.
	Attackers have costs to deploy attacks,
	and defenders incur costs to deploy defenses.
	{\it e.g.}, state-level agencies who pay big bucks
	to discover or buy 0-day exploits don't want
	to deploy them against ``low-value targets''
	and risk them getting discovered and patched.
	So while it may be infeasible to offer the general public
	strong protection against the best attacks by state-level adversaries,
	what we need to do (first) is raise the cost of attack
	to a point where low-value attacks don't work against most users,
	and adversaries are unwilling to deploy high-value attacks
	because of the risk of being discovered and neutralized.
}

\section{Introduction}\label{sec-introduction}

In today's ``Big Data'' Internet,
users often need to assume that, by default,
their every statement or action online
is monitored and tracked.
Users' statements and actions are routinely linked 
with detailed profiles 
built by entities ranging from commercial vendors and advertisers
to state surveillance agencies
to online stalkers and criminal organizations.  Indeed, recent events
have raised the stakes in Internet monitoring enormously.  
Documents leaked by Edward Snowden have revealed that the US government
is conducting warrantless surveillance on a massive scale and, in particular,
that the long-term goal of the National Security Agency is to be ``able
to collect virtually everything available in the digital 
world''~\cite{risen13nsa}.

Internet users often have legitimate need to be {\em anonymous} --
\ie, ``not named or identified'' by
\href{http://www.merriam-webster.com/dictionary/anonymous}{Webster's definition}
of the term --
to protect their online speech and activities
from being linked to their real-world identities.
Although the study of anonymous-communi\-ca\-tion technology
is often motivated by high-stakes use cases such as
battlefield communication, espionage, or political protest against
authoritarian regimes,
anonymity actually plays many well accepted roles in established 
democratic societies.  For example, paying cash, voting, opinion
polling, browsing printed material in a book store or library,
and displaying creativity and low-risk experimentalism
in forums such as \href{http://slashdot.org}{slashdot}
or \href{http://www.4chan.org}{4chan} are everyday examples of anonymous
activity.  Author JK Rowling used a pen name on a recent post-Harry 
Potter novel,
presumably not out of any fear of censorship or reprisal, but merely
``to publish without hype or expectation and \dots
to get feedback under a different name''~\cite{watts13rowling}.

Obtaining and maintaining anonymity on the Internet is challenging, however.
The state of the art in deployed tools,
such as Tor~\cite{torproject},
uses {\em onion routing} (OR) to relay encrypted connections
on a detour passing through randomly chosen relays
scattered around the Internet.  OR is scalable,
supports general-purpose point-to-point communication,
and appears to be effective against many of the attacks
currently known to be in use~\cite{gellman13secret}.
Unfortunately, OR is known to be vulnerable
to several classes of attacks
for which no solution is known or believed to be forthcoming soon.
For example, via {\em traffic confirmation},
an attacker who compromises a major ISP or Internet exchange
might in principle de-anonymize many Tor users in 
a matter of days~\cite{johnson13users}.
Through {\em intersection attacks}, an adversary can rapidly narrow
the anonymity of a target via actions linkable across time,
in much the same way
\href{https://www.aclu.org/blog/technology-and-liberty-national-security/surveillance-and-security-lessons-petraeus-scandal}{Paula Broadwell}
and the
\href{http://arstechnica.com/tech-policy/2013/08/how-cell-tower-dumps-caught-the-high-country-bandits-and-why-it-matters/}{``High Country Bandits''}
were de-anonymized~\cite{segal14catching}.
Finally, through software exploits or user error,
an attacker can often
\href{http://www.wired.com/threatlevel/2013/08/freedom-hosting/}{circumvent anonymity tools entirely}~\cite{wolinsky14managing}.

Current approaches to anonymity also appear unable to offer
accurate, principled measurement of the level or quality
of anonymity a user might obtain.
Considerable theoretical work
analyzes onion routing~\cite{feigenbaum12probabilistic},
but relies on idealized formal models
making assumptions that are unenforceable and may be untrue in real systems --
such as that users choose relays and communication partners at random --
or depending on parameters that are unknown in practice,
such as probability distributions representing user behavior.

We believe the vulnerabilities and measurability limitations
of onion routing
may stem from an attempt
to achieve an impossible set of goals
and to defend an ultimately indefensible position.
Current tools
offer a general-purpose, unconstrained,
and {\em individualistic} form of anonymous Internet access.
However, there are many ways for unconstrained, individualistic
uses of the Internet to be fingerprinted and tied to individual users.
We suspect that the only way to achieve 
measurable and provable levels of anonymity,
and to stake out a position defensible in the long term,
is to develop more {\em collective} anonymity protocols and tools.
It may be necessary to constrain the normally individualistic behaviors
of participating nodes, the expectations of users,
and possibly the set of applications and usage models
to which these protocols and tools apply.

Toward this end, we offer a high-level view of the Dissent project,
a ``clean-slate'' effort to build practical anonymity systems
embodying a collective model for anonymous communication.
Dissent's collective approach to anonymity is not and may never be
a ``drop-in'' functional replacement for Tor
or the individualistic, point-to-point onion routing model it implements.
Instead, Dissent sets out to explore radically different territory
in the anonymous\-/communication design space,
an approach that presents advantages, disadvantages,
and many as-yet-unanswered questions.
\com{
such as that certain well defined anonymity metrics
are more readily measurable and provable,
and disadvantages,
such as that 
}
An advantage is that the collective approach
makes it easier to design protocols
that provably guarantee certain well defined anonymity metrics
under arguably realistic environmental assumptions.
A disadvantage is that the collective approach
is most readily applicable to multicast-oriented communication,
and currently much less efficient or scalable than OR
for point-to-point communication.

Dissent follows in the tradition of Herbivore~\cite{sirer04eluding},
the first attempt to
build provable anonymity guarantees into a practical system,
and to employ dining cryptographers or DC-nets~\cite{chaum88dining}.
Dissent utilizes both DC-nets and 
verifiable shuffles~\cite{neff01verifiable},
showing for the first time how to scale the formal guarantees
embodied in these techniques to offer measurable anonymity sets
on the order of thousands of participants~\cite{wolinsky12dissent}.
Dissent's methods of scaling individual anonymity sets
are complementary and synergistic with techniques Herbivore pioneered
for managing and subdividing large peer-to-peer anonymity networks;
combining these approaches could enable further scalability improvements
in the future.

Dissent incorporates the first systematic countermeasures
to major classes of known attacks,
such as global traffic analysis
and intersection attacks~\cite{mathewson04disclosure,wolinsky13buddies}.
Because anonymity protocols alone cannot address
risks such as software exploits or accidental self\-/identification,
the Dissent project also includes Nymix,
a prototype operating system that hardens
the user's computing platform
against such attacks~\cite{wolinsky14managing}.
Dissent and Nymix OS can of course offer
only network-level anonymity, in which
the act of communicating does not reveal
which user sent which message.
No anonymity system can offer users {\em personal anonymity}
if, for example, they disclose their real-world identities
in their message content.

While at this time Dissent is still a research prototype
not yet ready for widespread deployment,
and may never be a direct replacement
for OR tools such as Tor
because of possibly fundamental tradeoffs,
we hope that it will increase the diversity of 
practical approaches and tools available for obtaining anonymity online.

\com{	subsumed by sentence I added above
This paper is concerned with {\it network-layer anonymity}.  
That is, the receiver of a message delivered by 
the anonymous-com\-mu\-ni\-ca\-tion
protocols that we consider does not know which network node is the
source of that message.  In Internet terms, this is {\it IP-layer anonymity}, 
meaning that the receiver would not know the IP
address (or perhaps even the domain) of the source.
This alone does not guarantee {\it personal anonymity}; the content of
the message may identify the human user who sent it even if the anonymity
system hides that user's network location.  The questions of whether and
how to anonymize content are orthogonal to the network-layer anonymity question,
and we do not address them in this paper.
}

Section~\ref{sec-onion} presents the basics of OR and Tor.  In
Section~\ref{sec:attacks}, we describe four problems with OR that have
gone unsolved for many years and may unfortunately be unsolvable.
Section~\ref{sec:dissent} provides an overview of the Dissent approach
to anonymous communication, and Section~\ref{sec:open} contains open
problems and future directions.

\section{Onion Routing and Tor}\label{sec-onion}

Currently the most widely deployed, general-purpose system for anonymous 
Internet communication is Tor~\cite{torproject}.
Tor's technical
foundation is onion routing~\cite{onion-routing:ih96},
derived in turn from mixnets~\cite{danezis03mixminion}.

Onion routing (OR) uses successive layers of encryption to route messages
through an overlay network, such that each node knows the
previous and the next node in the route but nothing else.  More 
precisely, let $(V,E)$ be a connected, undirected network and $R\subseteq V$ be
a set of nodes serving as {\it relays}. The set $R$ is known to all nodes
in $V$, as is the public key $K_r$, usable in some globally agreed-upon
public-key cryptosystem, for each node $r\in R$. 
There is a routing protocol that
any node in $V$ can use to send a message to any other node, but the nodes
need not know the topology $(V,E)$.  

\begin{figure}
\centering
\includegraphics[width=0.45\textwidth]{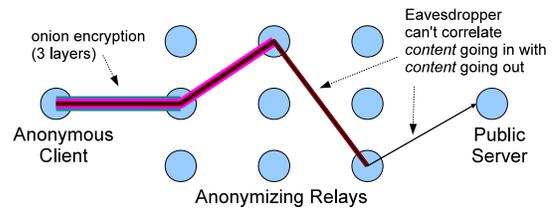}
\caption{Onion routing (OR).}
\label{fig:OR}
\end{figure}

If node $s$ wishes to send message $M$ to node $d$ anonymously,
$s$ first chooses a sequence $(r_1, r_2, \ldots, r_n)$ of relays.  
It then constructs an ``onion'' whose $n$ layers contain
both the message and the routing information needed to deliver it without
revealing node $s$'s identity to any node except the first relay $r_1$.  The 
core of the onion is $(d, M)$, {\it i.e.,} the destination node and
the message itself.
The $n$\textsuperscript{th} or innermost layer of the onion is
\[
O_n = (r_n, \enc{K_{r_n}}{d, M}),
\] 
{\it i.e.}, the $n$\textsuperscript{th} relay node and the encryption of the 
core under the $n$\textsuperscript{th} relay's public key.
More generally, the $i$\textsuperscript{th} layer $O_i$, $1\leq i\leq k-1$, is formed by
encrypting the $(i+1)$\textsuperscript{st} layer under the public key of the $i$\textsuperscript{th} relay
and then prepending the $i$\textsuperscript{th} relay's identity $r_i$:
\[
O_i = (r_i, \enc{K_{r_i}}{O_{i+1}}).
\]
Once it has finished constructing the outermost layer
\[
O_1 = (r_1,\enc{K_{r_1}}{O_2}),
\]
node $s$ sends $\enc{K_{r_1}}{O_2}$ to $r_1$, using
the routing protocol of the underlay network $(V,E)$.  When relay $r_i$,
$1\leq i\leq n$, receives $\enc{K_{r_i}}{O_{i+1}}$, it decrypts it using the 
private key
$k_{r_i}$ corresponding to $K_{r_i}$, thus obtaining both the identity
of the next node in the
route and the message that it needs to send to this next node (which it sends
using the underlying routing protocol). When $i=n$, the message is just the
core $(d,M)$, because, strictly speaking, there is no $O_{n+1}$.  We assume
that $d$ can infer from routing-protocol ``header fields'' of $M$ that it 
is the intended recipient and need not decrypt and forward. 
See Figure~\ref{fig:OR}.

\com{
Note that OR provides {\it network-layer anonymity}.  That is, the receiver of
traffic delivered by an OR overlay does not know which network node is the
source of that traffic.  In Internet terms, this is {\it IP-layer anonymity}, 
meaning that the receiver would not know the IP
address (or perhaps even the domain) of the source, but OR is a general
approach that can be used in any network, not just the Internet.
Note that this does not guarantee {\it personal anonymity}; the content of
the traffic may identify the human user who sent it even if the routing
information hides that user's network location.  The questions of whether and
how to anonymize content are orthogonal to the network-layer anonymity question,
and we do not address them in this paper.
}

Tor is a popular free-software suite based on OR. 
As explained on the Torproject 
website~\cite{torproject}, ``Tor protects you by bouncing your communications around 
a distributed network of relays run by volunteers all around the world; it 
prevents somebody watching your Internet connection from learning what sites
you visit, and it prevents the sites you visit from learning your [network] 
location.'' The project provides free application software that can be used for 
web browsing, email, instant messaging, Internet relay chat, file transfer,
and other common Internet activities; users can also obtain free downloads 
that integrate the underlying Tor protocol with established browsers,
email clients, {\it etc.}  Importantly, Tor users can easily (but are not
required to) transform their Tor installations into Tor relays, thus 
contributing to the overall capacity of the Tor network.  Currently, there are
approximately 40M ``mean daily users'' of Tor worldwide, slightly over 10\% of
whom are in the United States, and approximately 4700 relays.  These and other
statistics are regularly updated on the Tor Metrics Portal~\cite{tormetrics}.  

The IP addresses of Tor relays are listed in a public directory so that Tor
clients can find them when building circuits.  (Tor refers to routes as
``circuits,'' presumably because Tor is typically used for web browsing and
other TCP-based applications in which traffic flows in both directions between
the endpoints.)  Clearly, this makes it possible
for a network operator to prevent its users from accessing Tor.
The operator can
simply disconnect the first hop in a circuit, {\it i.e.}, the connection 
between the client and the first Tor relay, because the former
is inside the network and the latter is outside; this forces the Tor traffic to
flow through a network gateway, at which the operator can block it.  Several
countries that operate national networks, including China and Iran, have 
blocked Tor in precisely this way.  Similarly, website operators can block
Tor users simply by refusing connections from the last relay in a Tor circuit;
Craigslist is an example of a US-based website that does so.
As a partial solution,
the Tor project supports {\it bridges}, or relays whose IP
addresses are not listed in the public directory, of which there are currently
approximately 2000.  Tor bridges are just one of several anti-blocking
or {\it censorship-circumvention} technologies. 

There is inherent tension in OR between low latency, 
one aspect of which is short routes (or, equivalently, low values of $k$), 
and strong anonymity.  
Because its goal is to be a low-latency anonymous-communication 
mechanism, usable in interactive, real-time applications, Tor uses 3-layer
onions, {\it i.e.}, sets $k=3$ as in Figure~\ref{fig:OR}.  Despite this 
choice of small $k$, many potential users reject Tor 
because of its performance impact~\cite{dingledine09performance}.

\section{Attacks on Onion Routing}
\label{sec:attacks}

We now summarize four categories of known attacks to which OR is
vulnerable and for which no general defenses are known.

\paragraph{Global traffic analysis}

OR was designed to be secure against a {\em local adversary}, {\it i.e.},
one that might eavesdrop on some network links
and/or compromise some relay nodes
but only a small percentage of each.
It was not designed for security against traffic analysis
by a {\em global adversary} that can monitor large portions 
of the network constantly.

\begin{figure}
\centering
\includegraphics[width=0.45\textwidth]{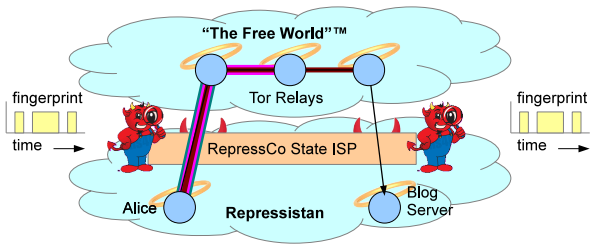}
\caption{Traffic confirmation or fingerprinting
	to de-anonymize onion-routing circuits}
\label{fig:trafconf}
\end{figure}

The most well known global-traffic-analysis attack---%
\emph{traffic confirmation}---%
was understood by Tor's designers
but considered an unrealistically strong attack model
and too costly to defend against \cite{torproject}.
In the standard scenario illustrated in Figure~\ref{fig:trafconf},
we assume that the attacker cannot break Tor's encryption
but can monitor both the encrypted traffic flowing from the user
to the first or {\em entry} relay
and the traffic flowing from the
final or {\em exit} relay to the user's communication partner.
This situation, while unlikely a decade ago, might be realistic today 
if both the user and her communication target are located in a 
single country, and the attacker is an ISP controlled or compromised
by a state-level surveillance agency.
In this case, the attacker in principle need only 
monitor the entry and exit traffic streams
and correlate them via known fingerprinting methods.

For decades, this {\it global-passive-adversary} attack model was
regarded as unrealistically strong,
and used to justify ``conservative'' assumptions
in formal models~\cite{feigenbaum12probabilistic}.
Unfortunately,
this adversarial model is now not only realistic but in fact too weak.
With the commercialization and widespread deployment
of routers that can perform deep packet inspection and modification,
including ``Man-in-the-Middle attacks''
against encrypted SSL streams at line rate \cite{gallagher13snowden},
it has become clear that any realistic adversary must be assumed
to be active, {\it i.e.}, able to modify traffic streams at will.

\paragraph{Active attacks}

\baf{mention/discuss FOXACID and man-on-the-side attacks?}

\begin{figure}
\centering
\includegraphics[width=0.45\textwidth]{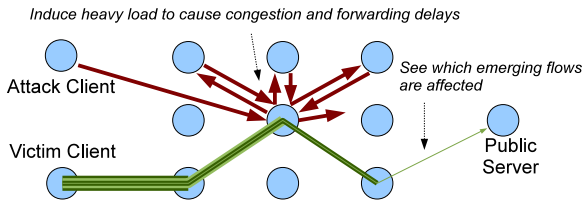}
\caption{Example of a congestion-based active attack}
\label{fig:congattack}
\end{figure}

The ability for an attacker to interfere actively in an anonymity network
creates a wide array of new attacks
as well as ways to strengthen existing traffic-analysis attacks.
Figure~\ref{fig:congattack} illustrates one example of
a {\em congestion attack}~\cite{evans09practical}.
In this scenario, we assume that the attacker can directly monitor
only one hop of a Tor circuit, {\it e.g.},
the traffic from the exit relay to the target web server.
The attacker in this case might be ``in the network''
or might simply own or have compromised the web server.
The attacker wishes to determine the set of relays
through which a long-lived circuit owned by a particular user passes.

The attacker chooses one relay at a time from Tor's public da\-ta\-base
and remotely attempts to increase that relay's load
by congesting it.  For example,
the attacker might simulate many ordinary Tor users
to launch a denial-of-service attack on the relay.
The attacker can amplify his power
by creating artificially long ``flower-petal'' circuits
that visit the target relay multiple times,
each visit interspersed with a visit to another relay,
as shown in Figure~\ref{fig:congattack}.
Regardless of how congestion is incurred,
it slows all circuits passing through this relay,
including the victim circuit, if and only if that circuit
passes through the targeted relay.
The attacker can therefore test whether a particular victim circuit
flows through a particular router,
simply by checking whether the victim circuit's average throughput
(which can be measured at any point along the circuit)
slows down during the period of attacker-generated congestion.
The attacker repeatedly probes different relays this way
until he identifies the victim's entry and middle relays.
Finally, the attacker might fully de-anonymize the user
by focusing traffic analysis on, or hacking,
the user's entry relay.

\com{
\paragraph{Denial-of-Security Attacks}

Another broad attack vector is to leverage denial-of-service (DoS) attacks
to compromise anonymity,
effectively strengthening conventional DoS attacks
into ``denial-of-security'' (DoSec) attacks.
In a classic scenario we call ``Borisov's Ambush''~\cite{borisov07denial},
we assume the attacker has infiltrated an OR network
with a substantial number of compromised relays,
for example by a large-scale hacking attack or a Sybil attack.
For any Tor circuit a victim attempts to route
through any attacker-controlled relay,
the attacker {\em selectively} blocks communication over that relay
only if the attacker {\em cannot} de-anonymize the circuit via
the traffic confirmation attack above:
{\it i.e.}, the attacker blocks circuits on which he controls
at least one relay but {\em not} both the entry and exit relays.
The attacker in this way forces the victim to ``reroll the dice''
by choosing another circuit at random;
although this reroll gives the victim another chance to get a circuit
containing {\em no} compromised node,
it also gives the {\em attacker} another chance to get
{\em two} compromised relays (entry and exit) on the victim's circuit
and hence to de-anonymize the circuit completely --
at which point the attacker will likely disable the DoS attack
and offer good service on that de-anonymized circuit.
}

\paragraph{Intersection attacks}

\begin{figure}
\centering
\includegraphics[width=0.45\textwidth]{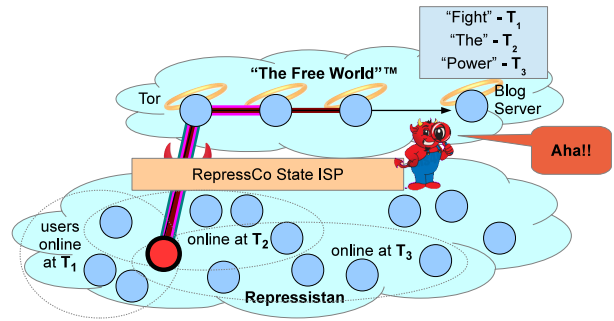}
\caption{Example of an intersection attack}
\label{fig:intersect}
\end{figure}

In most practical uses of anonymous communication,
a user typically needs to send not
just a single ``one-off'' message anonymously
but a sequence of messages that are explicitly related
and hence inherently linkable to each other.
For example, Tor clients need to maintain persistent TCP connections
and engage in back-and-forth ``conversations'' with web sites
in order to support interactive communication,
sending new HTTP requests that depend on the web server's
responses to the client's previous HTTP requests.
It is manifestly obvious at least to the web server
(and probably to any eavesdropper who can monitor the connection
between the Tor exit relay and the web site) 
which packets comprise the same Web communication session,
even if it is not (yet) clear who initiated that session.
Further, if the user leaves an anonymous browser window open
for an extended period
or regularly logs into the same anonymous Web mail account,
an eavesdropper may be able to link
many of the user's browsing sessions together over a long period of time.
Even if each message gives the attacker only a small
and statistically uncertain amount of information
just slightly narrowing the identity of the anonymous user,
combining this information across many observation points at different times
rapidly strengthens the attacker's knowledge.

In one example of this attack illustrated in Figure~\ref{fig:intersect},
an authoritarian government compels its ISPs or cellular carriers
to turn over logs of which customers were online
and actively using the network during which periods of time.
An anonymous dissident posts blog entries to a pseudonymous blog
at different points in time.
Assume that the attacker controls none of the user's onion relays.
Nor does he control the blog server; 
he merely observes the times at which the blog entries appeared
and the fact that the posts are manifestly linkable to each other,
and he can correlate this information with the ISP logs.
Perhaps the subject of the blog is official corruption in a particular city,
enabling the authoritarian state to guess that the dissident lives in that city
and narrow attention to a small set of local ISPs.
The attacker merely retrieves the sets of users who were online
at each time a blog post appeared and intersects those sets. 
Although there may be many thousands of users online
at each of these posting times individually,
all users other than the dissident in question
are likely to have gone offline during at least one of these times
(because of normal {\em churn} -- the partly random comings and goings 
of most users),
allowing the attacker to eliminate them from the victim's anonymity set.
The attacker simply needs to ``wait and watch'' until the dissident
has posted enough blog entries, and the intersection
of the online-user sets will shrink to a singleton.

The strength of this attack in practice is amply demonstrated by the fact
that similar reasoning is used regularly
in law enforcement~\cite{segal14catching}.
The FBI caught
\href{http://nakedsecurity.sophos.com/2013/12/20/use-of-tor-pointed-fbi-to-harvard-university-bomb-hoax-suspect/}{a Harvard student who used Tor to post a bomb threat}
by effectively intersecting the sets of Tor users
and Harvard network users at the relevant time.
\href{https://www.aclu.org/blog/technology-and-liberty-national-security/surveillance-and-security-lessons-petraeus-scandal}{Paula Broadwell was de-anonymized} via
the equivalent of an intersection attack,
as were
\href{http://arstechnica.com/tech-policy/2013/08/how-cell-tower-dumps-caught-the-high-country-bandits-and-why-it-matters/}{the ``High Country Bandits''}.
Intersection attacks also form the foundation
of \href{http://apps.washingtonpost.com/g/page/national/how-the-nsa-is-tracking-people-right-now/634/}{the NSA's CO-TRAVELER program},
which links known surveillance targets with unknown potential targets
as their respective cellphones move together from one cell tower
to another.

\baf{	Yet another great recent example Henry found: the Harvard bomb hoax.
	http://cbsboston.files.wordpress.com/2013/12/kimeldoharvard.pdf
}

\paragraph{Software exploits and self-identification}

\begin{figure}
\centering
\includegraphics[width=0.45\textwidth]{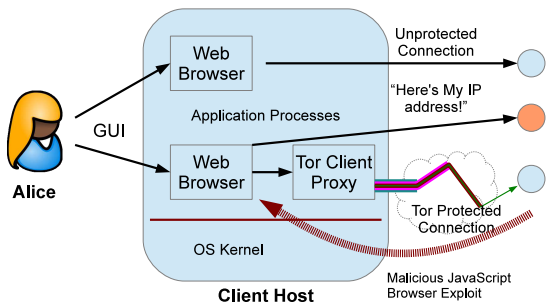}
\caption{Example of a software-exploit attack}
\label{fig:exploit}
\end{figure}

No anonymous communication system can succeed
if other software the user is running gives away his network location.
In a recent attack against the Tor network,
illustrated in Figure~\ref{fig:exploit},
a number of {\em hidden services} 
(web sites whose locations are protected by Tor
and which can be accessed only via Tor) 
were compromised so as to send malicious JavaScript code
to all Tor clients who connected to them.
This malicious JavaScript exploited a vulnerability
in a particular version of Firefox distributed as part of
the Tor Browser Bundle.
This exploit effectively ``broke out''
of the usual JavaScript sandbox and ran native code
as part of the browser's process.
This native code simply invoked the host operating system
to learn the client's true (de-anonymized) IP address, MAC address, 
{\it etc.}, and sent them
to an attacker-controlled server.

\com{
\paragraph{Accidental self-identification}

Finally, even if user's client software is uncompromised,
users can still accidentally ``give away'' their identities in myriad ways.
For example, many users are unaware that JPEG images
taken with common digital cameras
typically contain personally identifying information
such as the camera's make, model, and serial number,
the exact time and GPS coordinates at which the photo was taken, etc.
\baf{There are several amusing incidents; summarize/cite one?}
Even a user's writing style in a purely text-based anonymous 
chat or blogging forum might reveal his identity
to stylometric analysis~\cite{XXX}.
}

\section{Collective Anonymity in Dissent}\label{sec:dissent}

\com{
\baf{outline/notes:

Dissent architecture.  Principles:

1. Reliance on parallel DC-nets rather than serial relaying
	as anonymity foundation.

2. Collective network control plane.
	All decisions affecting what gets sent on the network, when,
	get determined by policy logic that has no access to
	sensitive/private data.

3. Conservative adversary modeling and anonymity metering
	for intersection attack and statistical disclosure protection,
	with active mitigation of anonymity loss.

Summarize key technical challenges:
- scalability.
- approaches to disruption resistance
(traps, Dissent v1, Dissent v2, Verdict).
- Robustness against server failure.
etc.

Lessons learned so far from Dissent project.

Performance prospects.
- for "appropriate" multicast-oriented applications.
- for web browsing, {\it e.g.}, WiNon/PriFi
}
}


\baf{Explain ``group anonymity'' and explicit anonymity sets
	as a basis for anonymity metrics, here or elsewhere earlier?}

As a step toward addressing these challenges,
we now introduce Dissent,
a project that expands the design space and
explores starkly contrasting foundations for anonymous communication.

\subsection{Alternative foundations for anonymity}

\com{
The security properties of mix-nets and onion routing
have been studied under a wide variety of formal models and simulated workloads.
The broad conclusion we draw from this body of work, unfortunately,
is that for unconstrained uses of these approaches---%
where each client or message sender makes independent, {\em individual}
choices of paths and message workloads---%
is ultimately insecure against strong traffic analysis.
\baf{move this to attacks section?}
}

Quantification and formal analysis of OR security
under realistic conditions has proven
an elusive goal~\cite{feigenbaum12probabilistic}.
Dissent therefore builds on alternative anonymity primitives
with more readily provable properties:
verifiable shuffles and dining cryptographers.

\paragraph{Verifiable shuffles} 

\baf{In retrospect this whole subsection needs to become
	no more than 1 paragraph.}

In a typical {\em cryptographic shuffle},
participating nodes play two disjoint roles:
there is a set of $n$ {\em clients} with messages to send
and a set of $m$ {\em shufflers} that randomly permute those messages.
Communication proceeds in synchronous rounds.
In each round, each of the $n$ clients encrypts a single message
under $m$ concentric layers of public-key encryption,
using each of the $m$ shufflers' public keys, in a standardized order.
All $n$ clients send their ciphertexts to the first shuffler,
which holds the private key to the outermost layer of encryption
in all the clients' ciphertexts.
The first shuffler waits until it receives all $n$ clients' ciphertexts,
then unwraps this outermost encryption layer,
randomly permutes the entire set of ciphertexts,
and forwards the permuted batch of $n$ ciphertexts to the next shuffler.
Each shuffler in turn unwraps another layer of encryption,
permutes the batch of ciphertexts, then forwards them to the next shuffler.
The final shuffler then broadcasts all the fully decrypted cleartexts
to all potentially interested recipients.

In an ``honest-but-curious'' security model
in which we assume each shuffler correctly follows the protocol
(without, for example, inserting, removing, or modifying any ciphertexts), 
the output from the last shuffler offers provable anonymity
among all non-colluding  clients, 
provided at least one of the shufflers keeps its random permutation secret.
Unfortunately, if any of the shufflers is actively dishonest,
this anonymity is easily broken.
For example, if the first shuffler duplicates the ciphertext
of some attacker-chosen client,
then the attacker may be able to distinguish the victim's cleartext
in the shuffle's final output simply by looking
for the cleartext that appears twice in the otherwise-anonymized output batch.

A substantial body of work addresses these vulnerabilities to such active 
attacks.
In a {\em sender-verifiable}
shuffle~\cite{brickell06efficient,corrigangibbs13proactively},
each client inspects the shuffle's output 
to ensure that its own message was not dropped, modified, or duplicated
before allowing the shuffled messages to be fully decrypted and used.
More sophisticated and complex {\em provable} shuffles,
such as Neff's~\cite{neff01verifiable},
enable each shuffler to prove to all observers the correctness
of its entire shuffle,
{\it i.e.}, that the shuffler's output is a correct permutation of its input,
without revealing any information about which permutation it chose.

\com{
in which each client first wraps its message in $2m$ encryption layers---%
$m$ {\em inner} layers followed by $n$ {\em outer} layers---%
before submitting the ciphertext to the first shuffler.
The $m$ shufflers then unwrap the $m$ outer layers while shuffling,
and submit the $n$ ``half-unwrapped'' ciphertexts to all clients for inspection.
Each client finally checks to ensure {\em its own} intermediate ciphertext
was preserved in the final shuffle,
before ``approving'' the shuffle by releasing the private key
needed to ...
}

Both types of verifiable shuffles offer cryptographic guarantees that
the process of shuffling reveals no information
about which of the $n$ clients submitted a given message
appearing in the shuffled output.
\com{	The introduction placed this issue "out of scope" of this paper,
	so it seems inappropriate to cite it as a disadvantage here,
	which might give the mistaken impression that this content problem
	is somehow specific to verifiable shuffles. -baf
This guarantee, unfortunately, does not automatically extend
to the larger system in which such a shuffle might be used.
For example, a client might still de-anonymize itself
with some information in the content of its message
or even by the client's presence or absence in a particular messaging round.
}
Shuffling has the practical disadvantage that the level of security
achievable against potentially compromised shufflers
depends on the number of shufflers in the path,
and multiple shufflers must inherently be placed in sequence
to improve security;
in essence, latency is inversely proportional to security.
The typical {\it cascade} arrangement above,
where all clients send their messages through the same
sequence of shufflers at the same time,
is most amenable to formal anonymity proofs,
but exacerbates the performance problem by
creating the ``worst possible congestion''
at each shuffler in succession
instead of randomly distributing load across many shufflers
as an ad hoc, individualistic OR network would.

For these reasons, verifiable shuffles may be practical
only when high latencies are tolerable,
and shufflers are well provisioned.
One relevant application is electronic voting,
for which some shuffle schemes were specifically intended,
and which might readily tolerate minutes or hours of latency.
A second application that arguably fits this model
is {\em anonymous remailers}~\cite{danezis03mixminion},
which were popular before onion routing.
Practical remailer systems have never to our knowledge
employed state-of-the-art verifiable shuffles
featuring anonymity proofs, however,
and were vulnerable to active attacks
analogous to the message duplication attack mentioned above.

\paragraph{Dining cryptographers}

The only well studied foundation for anonymity
not based on sequential relaying is
{\em Dining Cryptographers} or {\em DC-nets},
invented by David Chaum in the late 1980s~\cite{chaum88dining}
but never used in practical systems
until two decades later by Herbivore~\cite{sirer04eluding}.
Instead of relaying, DC-nets build on information-coding methods.

\begin{figure}
\centering
\includegraphics[width=0.40\textwidth]{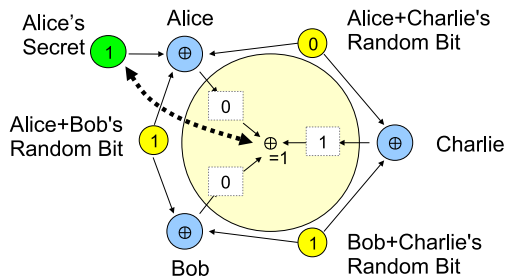}
\caption{The Dining Cryptographers approach to anonymous communication.
	Alice reveals a 1-bit secret to the group,
	but neither Bob nor Charlie learn which of the other two members
	sent this message.}
\label{fig:dc-nets}
\end{figure}

Consider Chaum's standard scenario, illustrated in Figure~\ref{fig:dc-nets}.
Three cryptographers are dining at a restaurant
when the waiter informs them that their meal has been paid for.
Growing suspicious,
they wish to learn whether one of their group paid the bill anonymously,
or NSA agents at the next table paid it.
So each adjacent pair of cryptographers flips a coin
that only the two can see.
Each cryptographer XORs the coins to his left and right
and writes the result on a napkin everyone can see---%
except any cryptographer who paid the bill (Alice in this case),
who flips the result of the XOR.
The cryptographers then XOR together the values written on all the napkins.
Because each coin toss affects the values of exactly two napkins,
the effects of the coins cancel out of the final result,
leaving a 1 if any cryptographer paid the bill (and lied about the XOR)
or a 0 if no cryptographer paid.
A 1 outcome provably reveals no information about
which cryptographer paid the bill, however:
Bob and Charlie cannot tell which of the other two cryptographers paid it
(unless of course they collude against Alice).

\begin{figure}
\centering
\includegraphics[width=0.30\textwidth]{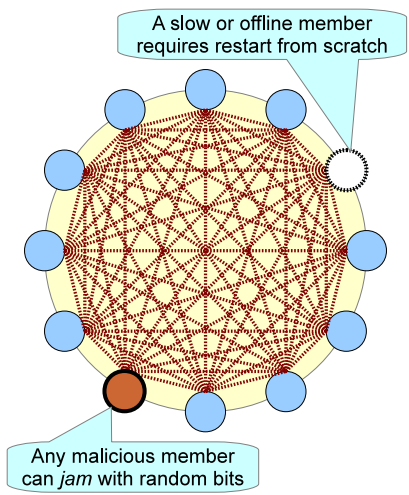}
\caption{Why DC-nets are hard to scale in practice:
	(1) worst-case $N \times N$ coin-sharing matrix;
	(2) network churn requires rounds to start over;
	(3) malicious members can anonymously jam the group.}
\label{fig:scaling}
\end{figure}

DC-nets generalize readily to support larger groups
and transmission of longer messages.
Typically each pair of cryptographers uses
Diffie-Hellman key exchange to agree on a shared seed
for a standard pseudorandom-bit generator,
which efficiently produces the many ``coin flips'' needed
to anonymize multi-bit messages.
While theoretically appealing, however,
DC-nets have not been perceived as practical,
for at least three reasons illustrated in Figure~\ref{fig:scaling}.
First, in groups of size $N$,
optimal security normally requires all pairs of cryptographers
to share coins, yielding complexity $\Omega(N^2)$, both computational and
communication.  Second, large networks of ``peer-to-peer'' clients
invariably exhibit high {\em churn},
with clients going offline at inopportune times;
if a DC-nets group member disappears during a round,
the results of the round become unusable and must be restarted from scratch.
Third, large groups are more likely to be infiltrated
by misbehaving members who might wish to block communication,
and any member of a basic DC-nets group can trivially---%
and anonymously---jam all communication
simply by transmitting a constant stream of random bits.

\baf{Also discuss the latency appeal of DC-nets:
	it's asymptotically no less efficient than dist-tree multicast,
	and latency doesn't depend on any security parameter.}

\subsection{Practical dining cryptographers}

Utilizing the DC-nets foundation in practical systems
requires solving two main challenges:
jamming and scalability.
Herbivore~\cite{sirer04eluding} pioneered the exploration of
practical solutions to both of these problems,
and the Dissent project continues this work.

\paragraph{The jamming problem}

Both Chaum's original paper~\cite{chaum88dining} and many follow-up works 
studied theoretical solutions to the jamming problem,
but were complex and to our knowledge never put into practice.
Herbivore sidestepped the jamming problem
by securely dividing a large peer-to-peer network
into many smaller DC-nets groups,
enabling a peer who finds himself in an unreliable or jammed group
to switch groups until he finds a functioning one.
This design has the advantage of scaling
to support arbitrary-sized networks,
with the downside that 
each peer obtains provable anonymity only within his own group --
typically tens of nodes at most --
and not guaranteeing anonymity within the larger network.
A second downside of switching groups to avoid jamming
is that 
an attacker who runs many Sybil nodes
and selectively jams only groups he cannot compromise completely,
while offering good service in groups in which he has isolated
a single ``victim'' node,
can make it more likely that a victim ``settles''
in a compromised group than an uncompromised one~\cite{borisov07denial}.

Dissent, the only system since Herbivore to put DC-nets into practice,
explores different solutions to these challenges.
First, Dissent addresses the jamming problem
by implementing {\em accountability} mechanisms,
allowing the group to revoke the anonymity of any peer 
found to be attempting to jam communication maliciously
while preserving strong anonymity protection for peers
who ``play by the rules.''  Dissent's first version
introduced a conceptually simple and clean accountability mechanism
that leveraged the verifiable-shuffle primitive discussed above,
at the cost of requiring a high-latency shuffle
between each round of (otherwise more efficient) DC-nets communication.
The next version~\cite{wolinsky12dissent}
introduced a more efficient but complex {\em retroactive-blame} mechanism,
allowing lower-latency DC-nets rounds to be performed ``back-to-back''
in the absence of jamming and requiring an expensive shuffle
only once per detected jamming attempt.

An adversary who manages to infiltrate a group
with many malicious nodes, however,
could still ``sacrifice'' them one-by-one
to create extended denial-of-service attacks.
Addressing this risk,
Dissent's most recent incarnation~\cite{corrigangibbs13proactively}
replaces the ``coins'' of classic DC-nets
with pseudorandom elliptic-curve group elements,
replaces the XOR combining operator with group multiplication,
and requires clients to prove their DC-nets ciphertexts correct
on submission,
using zero-knowledge proofs.
To avoid the costs of using elliptic-curve cryptography all the time,
Dissent implements
a hybrid mode that uses XOR-based DC-nets
unless jamming is detected, at which point the system switches to 
elliptic-curve DC-nets only briefly to enable the jamming victim
to broadcast an {\em accusation},
yielding a more efficient retroactive-blame mechanism.

\paragraph{Scaling and network churn}

\begin{figure}
\centering
\includegraphics[width=0.45\textwidth]{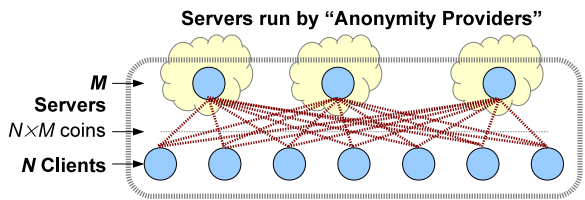}
\caption{Improving scalability and churn resistance through
	an asymmetric, client/server DC-nets architecture.}
\label{fig:anytrust}
\end{figure}

\baf{Make the figure more visually clear that the 3 servers
	might be chosen from a large cloud of anonymity providers,
	similar to Tor onion relays?}

Even with multiple realistic solutions to the jamming problem now available,
DC-nets cannot offer useful anonymity if they can guarantee anonymity-set sizes
of at most tens of members.
Herbivore addressed the $N \times N$ communication complexity problem
via a star topology,
in which a designated member of each group collects other members' ciphertexts,
XORs them together, and broadcasts the results to all members.
Without a general solution to the network churn and jamming problems,
however, both Herbivore and the first version of Dissent
were limited in practice to small anonymity sets 
comprising at most tens of nodes.

To address churn and scale DC-nets further,
Dissent now adopts a client/multi-server model
with trust split across several servers,
preferably administered independently.
No single server is trusted; in fact, Dissent preserves maximum security
provided only that not all of a group's servers maliciously collude
against their clients.
The clients need not know or guess which server is trustworthy
but must merely trust that at least one trustworthy server exists.

When a Dissent group is formed, the group creator
defines both the set of servers to support the group
and the client-admission policy; in the simplest case, the policy is 
simply a list of public keys representing group members.
Dissent servers thus play a role analogous to relays in Tor,
serving to support the anonymity needs of many different clients and groups.
Like Tor relays, the Dissent servers supporting a new group might 
be chosen automatically
from a public directory of available servers to balance load.
Choosing the servers for each group
from a larger ``cloud'' of available servers in this way
in principle enables Dissent's design to support an arbitrary number of groups,
but the degree to which an individual group scales
may be more limited.
If a particular logical group becomes extremely popular,
Herbivore's technique of splitting a large group
into multiple smaller groups may be applicable.
Our current Dissent prototype does not yet implement
either a directory service or Herbivore-style subdivision of large networks,
however.

While individual groups do not scale indefinitely,
Dissent exploits its client/multi-server architecture
to make groups scale two orders of magnitude
beyond prior DC-nets designs~\cite{wolinsky12dissent}.
As illustrated in Figure~\ref{fig:anytrust},
clients no longer share secret ``coins'' directly with other clients
but only with each of the group's servers.
Since the number of servers in each group is typically small
({\it e.g.}, 3--5, comparable to the number of Tor relays supporting a circuit),
the number of pseudorandom strings each client must compute
is substantially reduced.
This change does not reduce anonymity, however,
subject to Dissent's assumption that at least one server is honest.
Chaum's DC-nets security proof~\cite{chaum88dining} ensures ideal anonymity
provided all honest nodes are connected via the coin-sharing graph;
Dissent satisfies this requirement,
because the one honest server assumed to exist
shares coins directly with all honest clients.


More importantly in practice,
Dissent's client/multi-server coin-sharing design
addresses network churn by
making the composition of client ciphertexts independent of
the set of other clients online in a given round.
The servers set a deadline,
and all clients currently online must submit their ciphertexts
by that deadline or risk being ``left out'' of the round.
Unlike prior DC-nets designs,
if some Dissent clients miss the deadline,
the other clients' ciphertexts remain usable.
The servers merely adjust the set of
client/server-shared secrets they use to compute
their server-side DC-net ciphertexts.
Because each client's ciphertext depends on secrets
it shares with all servers,
no client's ciphertext can be used or decrypted
unless all servers agree on the same set of online clients in the round
and produce correct server-side ciphertexts based on that agreement.
Malicious servers can at most corrupt a round
and cannot de-anonymize clients except by colluding with all other servers.

\baf{	Discuss how the Dissent group communication primitive
	scales for multicast purposes,
	and how it gets used in applications?}

\subsection{How Dissent handles attacks}

We now summarize how Dissent handles the attacks
in Section~\ref{sec:attacks}.

\paragraph{Global traffic analysis}

Dissent builds on anonymity primitives
that have formal security proofs
in a model where the attacker is assumed to monitor
all network traffic sent among all participating nodes
but cannot break the encryption.
We have extended these formal security proofs
to cover the first version of the
full Dissent protocol~\cite{syta14security},
and formal analysis of subsequent versions is in progress.
Although verifiable shuffles differ from DC-nets in their details,
both approaches share one key property that enables formal anonymity proofs:
All participants act collectively
under a common ``control plane'' rather than individually as in 
an ad hoc OR system.
For example, they send identical amounts of network traffic in each round,
although amounts and allocations may vary from round to round.

\paragraph{Active attacks}

\begin{figure}
\centering
\includegraphics[width=0.45\textwidth]{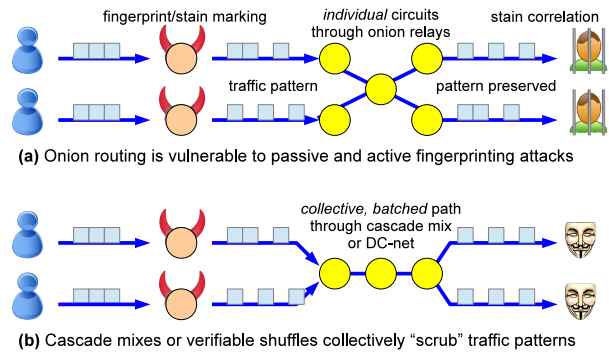}
\caption{Fingerprinting or staining attacks}
\label{fig:fingerprint}
\end{figure}

One countermeasure to traffic analysis in OR
is to ``pad'' connections to a common bit rate.
While padding may limit passive traffic analysis,
it often fails against active attacks,
for reasons illustrated in Figure~\ref{fig:fingerprint}.  Suppose
a set of OR users pad the traffic they send to a common rate,
but a compromised upstream ISP wishes to ``mark'' or ``stain''
each client's traffic by delaying packets with a distinctive timing pattern.
An OR network, which handles each client's circuit individually,
preserves this recognizable timing pattern (with some noise)
as it passes through the relays,
at which point the attacker might recognize the timing pattern at the egress
more readily than would be feasible with a traffic-confirmation attack alone.
\com{	cite below instead
Recent padding approaches add resistance to
certain active attacks~\cite{leblond13anon},
but no general or provable solution to active atacks in OR
currently appears forthcoming.
}
Active attacks also need not mark circuits solely via timing.
\href{http://blog.torproject.org/blog/tor-security-advisory-relay-early-traffic-confirmation-attack}{%
A sustained attack deployed against Tor last year}
exploited another subtle protocol side-channel to mark and correlate circuits,
going undetected for five months
before being discovered and thwarted last July.

The collective-anonymity primitives
underlying Herbivore and Dissent, in contrast,
structurally keep the clients comprising an anonymity set
in ``lock-step,''
under the direction of a common, collective control plane.
As in the popular children's game ``Simon Says,''
participants transmit when and how much
the collective control plane tells them to transmit.
A client's network-visible communication behavior
does not leave a trackable fingerprint or stain,
even under active attacks such as those above,
because its network-visible behavior depends {\em only}
on this anonymized, collective control state;
that is, a client's visible behavior never depends directly
on individual client state.
Further, the Dissent servers implementing this collective control plane
do not know which user owns which pseudonym or DC-nets transmission slot
and thus cannot leak that information via their decisions,
even accidentally.

Contrary to the intuition
that defense against global traffic analysis and active attacks
require padding traffic to a constant rate,
Dissent's control plane can adapt flow rates to client demand
by scheduling future rounds based on (public) results from prior rounds.
For example, the control-plane scheduler dynamically allocates
DC-nets transmission bandwidth to pseudonyms who
in prior rounds anonymously indicated a desire to transmit
and hence avoids wasting network bandwidth or computation effort
when no one has anything useful to say.
Aqua, a recent project to strengthen OR security,
employs a similar collective-control philosophy
to normalize flow rates dynamically
across an anonymity set~\cite{leblond13anon}.
In this way, a collective control plane can in principle
not only protect against both passive and active attacks
but, ironically, can also improve efficiency
over padding traffic to a constant bit rate.

\paragraph{Intersection attacks}

While the power and generality of intersection attacks has been
extensively studied in the past decade,
there has been scant work on actually building mechanisms
to protect users of practical systems
against intersection attacks.
The nearest precedents we are aware of are suggestions
that traffic padding may make intersection attacks
more difficult~\cite{mathewson04disclosure}.
To the best of our knowledge, such proposals have never been 
implemented, in part because there is no obvious way to measure
how much protection against intersection attacks
a given padding scheme will provide in a real environment.

Dissent is the first anonymity system designed with mechanisms
both to measure potential vulnerability to intersection attacks,
using formally grounded but plausibly realistic metrics,
and to offer users active control over anonymity loss
under intersection attacks~\cite{wolinsky13buddies}.
Dissent implements two different anonymity metrics:
{\em possinymity}, a possibilistic measurement of anonymity-set size
motivated by ``plau\-si\-ble-deniability'' arguments,
and {\em indinymity}, an indistinguishability metric
effective against stronger adversaries that may make
probabilistic ``guesses'' via
statistical disclosure~\cite{mathewson04disclosure}.

Users may set policies for long-lived pseudonyms
limiting the rate at which measured possinymity or indinymity may be lost,
or setting a threshold below which these metrics must not fall.
Dissent's collective control plane enforces these policies in essence
by detecting when allowing a communication round to proceed
might reduce a pseudonym's possinymity or indinymity ``too much''
and, in response, suppressing or delaying communication temporarily.
The control plane can compute these metrics and enforce these policies
even though its logic does not ``know'' which user actually owns each pseudonym.
The downside is that
employing these controls to resist intersection attacks
can reduce the 
responsiveness, availability, and/or lifetime of a pseudonym.
We believe this cost reflects a fundamental tradeoff
between anonymity and availability.

\paragraph{Software exploits and self-identification}

\begin{figure}
\centering
\includegraphics[width=0.45\textwidth]{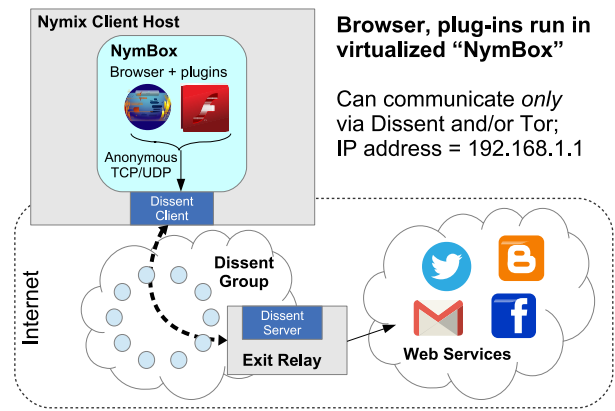}
\caption{Nymix: using per-pseudonym virtual machines or NymBoxes
	to harden the client operating system
	against software exploits, staining, and self-identification}
\label{fig:winon}
\end{figure}

No anonymity protocol, by itself, can prevent de-anonymization
via software exploits or user self-identification.
Nevertheless, the Dissent project
is exploring system-level solutions to this problem
via Nymix, a prototype USB-bootable Linux distribution
that employs virtual machines (VMs) to improve resistance
to exploits~\cite{wolinsky14managing}.

As shown in Figure~\ref{fig:winon},
Nymix runs anonymity-client software
(currently either Tor or Dissent) in the platform's host operating system
but isolates the browser and any plug-ins and other extensions it may depend on
in a separate Guest VM.
No software in this guest VM is given access to information
about the physical host OS or its network configuration.
For example, the guest VM sees only a standard private (NATted)
IP address such as 192.168.1.1 and the fake MAC address of a virtual device.
Even native code injected by the \href{http://www.wired.com/2014/08/operation_torpedo/}{recent Tor Browser Bundle exploit}
would thus not be able to ``leak'' the client's IP address
without also breaking 
out of the VM (which of course may be possible,
but raises the attack difficulty).

Nymix binds guest-VM state instances to pseudonyms
managed by the anonymity layer,
enabling users to launch multiple simultaneous pseudonyms
in different VMs or {\em NymBoxes}.
Nymix securely discards all pseudonym state embodied in a NymBox when desired
to minimize the user's long-term exposure to intersection attacks.
This binding of pseudonyms to VMs
makes it easy for the user to maintain state related to the context
of one logical pseudonym (such as Web cookies, open logins, {\it etc.}),
while offering stronger protection against the user's
accidentally linking different pseudonym VMs,
because they appear as entirely separate OS environments
and not just different browser windows or tabs.

To reduce the risk of self-identification,
Nymix allows the user to ``move'' data between non-anonymous contexts, 
such as personal JPEG photos stored on the host OS,
and pseudonym-VM contexts
only via a {\em quarantine} file system ``drop box.''
Any files the user moves across browsing contexts in this way
undergoes a suite of tests for possibly compromising information, 
such as EXIF metadata within JPEGs.
The quarantine system warns the user of any detected compromise risks
and gives him the opportunity
to scrub the file or decide not to transfer it at all.
While all of these defenses are inherently ``soft,'' because
there is only so much we can do to prevent users from
shooting themselves in the foot, 
Nymix combines these VM-based isolation and structuring principles
in an effort to make it easier for users to make 
appropriate and well informed uses of today's and tomorrow's anonymity tools.


\baf{Add the data-scrubbing dropbox to the winon figure?}

\section{Challenges and Future Work}\label{sec:open}

Dissent takes a few steps in developing
a collective approach to anonymous communication,
but many practical challenges remain.

First, while DC-nets now scale to thousands of users,
they need to scale to hundreds of thousands or more.
One approach is to combine Dissent's scaling techniques
with those of Herbivore~\cite{sirer04eluding}
by dividing large anonymity networks
into manageable anonymity sets ({\it e.g.}, hundreds or thousands of nodes),
balancing performance against anonymity guarantees.
A second approach is to use small, localized Dissent clusters,
which already offer performance adequate
for interactive Web browsing~\cite{wolinsky12dissent,wolinsky14managing},
as a decentralized implementation for the crucial entry-relay role
in a Tor circuit~\cite{torproject}.  Much of a Tor user's security
depends on his entry relay's being uncompromised~\cite{johnson13users};
replacing this single point of failure with a Dissent group
could distribute the user's trust among the members of this group
and further protect traffic between the user and the Tor relays
from traffic analysis by ``last mile'' ISP adversaries.

Second, while Dissent can measure vulnerability to intersection attack
and control anonymity loss~\cite{wolinsky13buddies},
it cannot also ensure availability
if users exhibit high churn and individualistic,
``every user for himself'' behavior.
Securing long-lived pseudonyms
may be feasible only in applications
that incentivize users to keep 
communication devices online consistently,
even if at low rates of activity,
to reduce anonymity decay caused by churn.
Further, robust intersection-attack resistance may be practical
only in applications designed to encourage users to act collectively,
rather than individually,
and optimized for these collective uses.

Applications in which users cooperatively
produce collective information ``feeds'' consumed by many others users
may be well suited to Dissent's collective anonymity model:
\eg, the interaction models of IRC,
forums like Twitter or Slashdot,
or applications supporting voting, deliberating, or ``town hall'' meetings.
Given the close relationship between collective deliberation
and the foundations of democracy and freedom of speech,
such applications may also represent some of the most socially important
use cases for online anonymity.
How best to support and incentivize cooperative 
behavior, however, remains an important open problem.

Finally, it is clear that large anonymity sets require widespread public
demand for anonymity.  Tor's 40M ``mean daily users'' are dwarfed in number
by the users of Google, Facebook, Yahoo!, and other services that do not
provide anonymity\ --\ and cannot provide it, because their business
models depend crucially on exploitation of personal information.  Public
demand for anonymity online may rise as a result of the ongoing
surveillance scandal, thereby providing an opportunity to deploy new
anonymity tools.


\begin{footnotesize}
\bibliographystyle{plain}
\bibliography{net,sec}
\end{footnotesize}

\end{document}